\documentclass[useAMS]{mn2e}
\usepackage{graphicx}

\def\simgt{\ {\raise-.5ex\hbox{$\buildrel>\over\sim$}}\ }

\begin{document}

\title[The pulsating DA white dwarf star EC\,14012$-$1446]
{The pulsating DA white dwarf star EC\,14012$-$1446: results from four 
epochs of time-resolved photometry}
\author[G. Handler et al.]
 {G. Handler,$^{1}$ E. Romero-Colmenero,$^{2}$ J. L. Provencal,$^{3}$ K. 
Sanchawala,$^{4, 5}$ \and M. A. Wood,$^{6}$ I. Silver,$^{6}$ W.-P. 
Chen\,$^{4}$
\and \\
$^{1}$ Institut f\"ur Astronomie, Universit\"at Wien,
T\"urkenschanzstrasse 17, A-1180 Wien, Austria\\
$^{2}$ South African Astronomical Observatory, P.O. Box 9, Observatory 
7935, South Africa\\
$^{3}$ Mt.\ Cuba Observatory \& Dept.\ of Physics and Astronomy, 
University of Delaware, 223 Sharp Laboratory, Newark, DE 19716\\
$^{4}$ Graduate Institute of Astronomy, National Central University, 
Chung-Li 32054, Taiwan\\
$^{5}$ Physics Department, Birla Institute of Technology and Science, 
Pilani 333 031, India\\
$^{6}$ Dept.~of Physics and Space Sciences \& SARA Observatory, Florida
Institute of Technology, Melbourne, FL 32901, USA}

\date{Accepted 2007 July 17.
 Received 2007 August 13;
in original form 2007 September 10} 
\maketitle 
\begin{abstract} 

The pulsating DA white dwarfs are the coolest degenerate stars that
undergo self-driven oscillations. Understanding their interior structure
will help to understand the previous evolution of the star. To this end,
we report the analysis of more than 200 h of time-resolved CCD photometry
of the pulsating DA white dwarf star EC~14012$-$1446 acquired during four
observing epochs in three different years, including a coordinated
three-site campaign. A total of 19 independent frequencies in the star's
light variations together with 148 combination signals up to fifth order
could be detected. We are unable to obtain the period spacing of the
normal modes and therefore a mass estimate of the star, but we infer a
fairly short rotation period of $0.61\pm0.03$\,d, assuming the
rotationally split modes are $\ell=1$. The pulsation modes of the star
undergo amplitude and frequency variations, in the sense that modes with
higher radial overtone show more pronounced variability and that amplitude
changes are always accompanied by frequency variations. Most of the
second-order combination frequencies detected have amplitudes that are a
function of their parent mode amplitudes, but we found a few cases of
possible resonantly excited modes. We point out the complications in the
analysis and interpretation of data sets of pulsating white dwarfs that
are affected by combination frequencies of the form $f_A+f_B-f_C$
intruding into the frequency range of the independent modes.

\end{abstract}


\section{Introduction}

White dwarf stars are the most common end point of stellar evolution.  
Almost all stars with masses below $8 M_{\odot}$ end their lives in the
white dwarf stage, and most of them do so after being the central star of
a Planetary Nebula. Once the central star has exhausted its nuclear fuel,
it slowly cools and dims at nearly constant radius. As there is a definite
low-temperature cutoff to the white dwarf luminosity function, it can be
used to determine the age of the galactic disk (e.g.\ Winget et al.\ 1987,
Richer et al.\ 2000) or stellar clusters (e.g.\ Hansen et al.\ 2007),
although some theoretical uncertainties remain (see Fontaine, Brassard \&
Bergeron 2001 for a recent review).

While white dwarf stars cool, they cross three pulsational instability
strips, i.e.\ regions where they exhibit self-driven nonradial
oscillations. The hottest instability region contains pulsating PG~1159
stars (the GW Vir stars), the intermediate one consists of pulsating DB
white dwarfs (the V777 Her stars), whereas pulsating DA stars (the ZZ Ceti
stars) constitute the coolest white dwarf instability strip.  If all white
dwarfs showed self-excited pulsations as they pass through their
instability domain, this would mean that the interior structure of the
pulsators is representative of all white dwarfs. The ZZ Ceti instability
strip is considered to be pure by most authors (e.g.\ see Bergeron et al.\
2004, Mukadam et al.\ 2004, Castanheira et al.\ 2007).

In any case, if a large number of oscillations can be detected and
resolved in a given star, one can apply asteroseismic methods to it. In
other words, the oscillations can be used to determine the interior
structure of such stars. This is important for astrophysics in general, as
the history of evolution of stars is engraved in the interiors of its
end product, the white dwarf star. Fortunately, white dwarfs are well
suited for asteroseismology because virtually complete pulsational mode
spectra over a certain frequency interval were found in a number of cases
(e.g. Winget et al.\ 1991, 1994), for instance for the prototype pulsating 
PG~1159 and DB white dwarf stars.

The pulsating DA white dwarfs have remained less accessible to this
method. Their interior structure is somewhat more complicated due to their
surface hydrogen layer and possible interior crystallization (e.g.\
Montgomery \& Winget 1999). Their pulsational mode spectra are in general
less dense than those of the pulsating PG~1159 and DB stars, and even so,
some of the observed signals are difficult to be reconciled with
theoretical models (e.g.\ Pech, Vauclair \& Dolez 2006, or Castanheira et
al.\ 2004).

Part of these difficulties may arise from combination frequencies, i.e.\
signals that may only reflect the Fourier decomposition of the
non-sinusoidal light curve shapes and that do not correspond to normal
mode oscillations, intruding into the frequency domain of the normal
modes. Another potential problem is the occurrence of amplitude and
frequency variations on short time scales (e.g.\ Handler et al.\ 2003),
resulting in spurious peaks that complicate the identification of normal
modes. On the other hand, the temporal variations of the pulsational
spectra of some ZZ Ceti stars may make modes that are not permanently
observable visible at some periods of observation. This fact was first
recognized and taken advantage of by Kleinman et al.\ (1998).

Still, considerable progress in the understanding of ZZ Ceti star
pulsation has been made recently. For instance, Kepler et al.\ (2005)  
pointed out how the C/O ratio in the core of these stars can be measured
from their evolutionary period changes that reflect their cooling rate.
Furthermore, based on the original idea by Brickhill (1992), Montgomery
(2005) derived a method to constrain pulsational mode identifications from
the light curve shapes of pulsating white dwarfs, and to recover the
thermal response time scale of the convection zone, which depends on
effective temperature.

Consequently, it is desirable to obtain extensive observations of
multiperiodic ZZ Ceti stars, and to decipher their pulsational mode
spectra. A good target for asteroseismic studies must be carefully chosen:  
in general, hot pulsating DA white dwarfs have few modes, low amplitudes
and short periods, whereas the cooler pulsators have longer periods,
higher amplitudes and more modes, but temporal variations in their
pulsation spectra. The best choice may then lie in between: when observing
pulsators in the centre of the ZZ Ceti instability strip, short and long
periods, high amplitudes, as well as stable mode spectra may be present.

\section{Observations and reductions}

The ZZ Ceti star EC~14012$-$1446 was discovered to be a multiperiodic
high-amplitude pulsator by Stobie et al.\ (1995), who found five
independent and five combination frequencies in their data. As this object
seemed well suited for asteroseismic purposes, we have obtained
time-resolved CCD photometry of the star over four epochs distributed over
three years. Three of the data sets were acquired at a single site, but
one data set was obtained during a coordinated campaign involving three
observatories well spread in geographical longitude.

The first measurements originated from the 0.75-m telescope at the South
African Astronomical Observatory (SAAO) and were taken in April 2004. A
high-speed CCD photometer (O'Donoghue 1995) was used in frame-transfer
mode with integration times of 10 seconds. No filter was employed.

Two months later, we observed EC~14012$-$1446 with the 1.9-m telescope at
the SAAO with the same photometer. Due to the smaller field of view,
frame-transfer mode was not employed, but the integration time of 10
seconds was kept, resulting in one frame per 12 seconds, and again no
photometric filter was inserted.

In the year 2005, we organized a multi-site campaign for the star,
involving three observatories in South Africa, the United States and
Taiwan. In this way, close-to uninterrupted measurements could be obtained
during the central part of the campaign. The 1.0-m telescope at Lulin
Observatory started the observations, which were taken as unfiltered
12-second integrations, resulting in one frame per 17 seconds. The 1.0-m
telescope at SAAO came on line four nights afterwards, this time using a
standard CCD with 21-second exposures and no filter, yielding one data
frame every 30 seconds. Later during the same night, the 0.9-m telescope
of the Southeastern Association for Research in Astronomy (SARA) located
in Arizona joined the campaign, acquiring unfiltered measurements with
12-second exposures for one data point per 16 seconds.

Finally, in 2007, the 0.9-m telescope at the Cerro Tololo Interamerican
Observatory (CTIO) was used to measure EC~14012$-$1446 again. These data
were acquired through a S8612 red-cutoff filter. The integration time was
10 seconds, leading to a data rate of one frame per 30 seconds. The star
was observed every second night of this run. An overview of all the
observing runs is given in Table~1.

\begin{table}
\caption[]{Overview of the observations. $\Delta$T is the total time of 
monitoring, and $\Delta$f is the frequency resolution of the data set.
For the multisite campaign, the frequency resolution of the combined
data is quoted and marked with an asterisk.}
\begin{center}
\begin{tabular}{llccc}
\hline
Month/Year & Telescope & $\#$ nights & $\Delta$T & $\Delta$f \\
& & & (hr) & ($\mu$Hz) \\
\hline
April 2004 & SAAO 0.75-m & 4 & 36.4 & 2.2\\
June 2004 & SAAO 1.9-m & 4 & 18.6 & 3.7\\
May 2005 & Lulin 1.0-m & 7 & 26.8 & 0.75$^\ast$\\
May 2005 & SAAO 1.0-m & 8 & 50.9 & 0.75$^\ast$\\
May 2005 & SARA 0.9-m & 5 & 30.5 & 0.75$^\ast$\\
April 2007 & CTIO 0.9-m & 7 & 40.1 & 0.87\\
\hline
Total & & 35 & 203.3\\
\hline
\end{tabular}
\end{center}
\end{table}

All measurements from the years 2004 and 2005 were reduced in the same
way. First, standard IRAF\footnote{IRAF, the Image Reduction and Analysis
Facility, is written and supported by the IRAF programming group at the
National Optical Astronomy Observatories (NOAO) in Tucson, Arizona.}
routines were used to correct the images for overscan, bias level (if
needed), dark counts (if needed) and flat field. Photometry was carried
out using the MOMF (Multi--Object Multi--Frame, Kjeldsen \& Frandsen 1992)  
package. MOMF applies combined Point--Spread Function/Aperture photometry
relative to an optimal sample of comparison stars, ensuring 
highest-quality differential light curves of the target.

The CTIO data from 2007 were also reduced with IRAF in the above-mentioned
way, but photometry was carried out with a series of IRAF scripts
employing aperture photometry optimized for high-speed CCD data (Kanaan,
Kepler \& Winget 2002). The final output from these scripts again is a
differential light curve of the target star. Tests showed that both of the
CCD photometry programs employed by us yield results of comparable
quality.

All differential light curves were visually inspected and data obtained
during poor photometric conditions were removed. As EC~14012$-$1446 is
substantially bluer than the nearby stars used as comparisons, some
differential colour extinction present in the light curves was carefully
removed with the Bouguer method (fitting a straight line to a plot of
magnitude vs.\ air mass). No low-frequency filtering was performed to
avoid affecting possible intrinsic long-period signals. Finally, as no
periodicities shorter than 140 seconds were present in the light curves,
some data were merged to speed up the computations to follow: the April
2004 data were co-added to 20-second bins, the June 2004 data to 24-second
bins, the Lulin measurements from 2005 to 34-second bins and the SARA data
from 2005 to 32-second bins. This procedure also ensured that the 2005
campaign data all have similar time sampling and therefore similar weight
in frequency analyses.

Finally, all measurements were transformed to a common time base. We chose
Terrestrial Time (TT) as our reference for measurements on the Earth's
surface and applied a correction to account for the Earth's motion around
the solar system's barycentre. As this barycentric correction varied by
up to $-1.7$~s during single observing nights, we applied it point by
point. Our final time base therefore is Barycentric Julian Ephemeris Date
(BJED). The resulting time series was subjected to frequency analysis. 

\section{Frequency analysis}

The frequency analysis was mainly performed with the program {\tt
Period98} (Sperl 1998). This package applies single-frequency power
spectrum analysis and simultaneous multi-frequency sine-wave fitting. It
also includes advanced options such as the calculation of optimal
light-curve fits for multi-periodic signals including harmonic and
combination frequencies. Our analysis will require these features.

We began by computing the Fourier spectral window functions of all data
sets, and show the results in Fig.\ 1. The multisite campaign (May 2005)
resulted in a reasonably clean window, whereas the other data sets suffer
from aliasing. The April 2004 window has strong sidelobes not only at the
daily alias 11.6\,$\mu$Hz away from the correct frequency, but also at
2.8, 8.8 and 14.3\,$\mu$Hz, respectively. The June 2004 measurements have
the poorest window function with low frequency resolution and strong
diurnal aliases, and the April 2007 data suffer not only from daily
aliasing, but also from spurious signals 5.2, 6.4, 16.9 and 18.0\,$\mu$Hz
away from the correct frequency. This has to be kept in mind when carrying
out a frequency search.

\begin{figure}
\includegraphics[width=88mm,viewport=00 00 260 355]{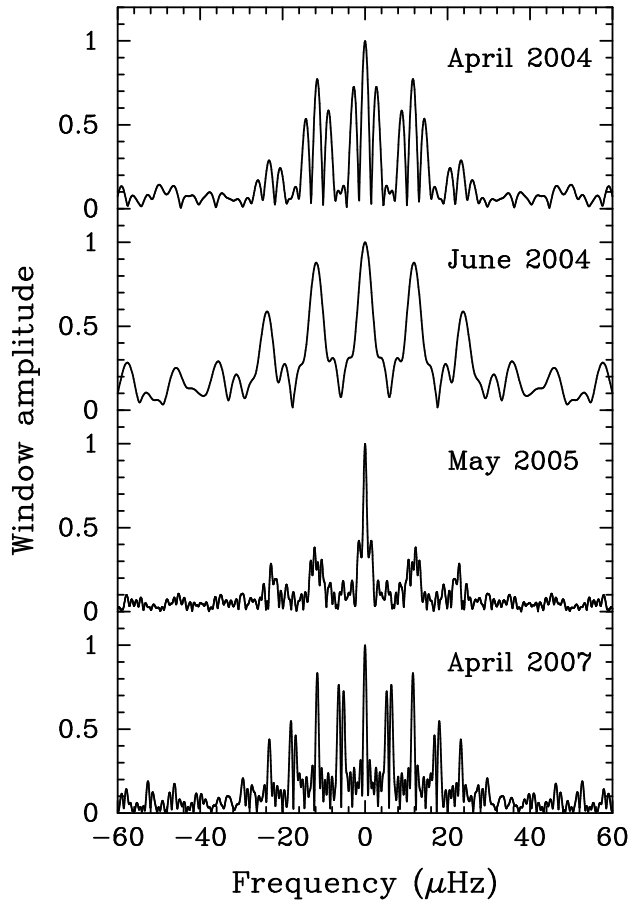}
\caption[]{The spectral window functions of our data sets from the four 
different epochs.}
\end{figure}

We proceeded by computing the Fourier amplitude spectra of the individual
data sets, which can be found on the right-hand side of Fig.\ 2. Both the
light curves on the left-hand side of Fig.\ 2 as well as the amplitude
spectra make it immediately obvious that the dominant time scale of the
light variations of EC~14012$-$1446 increased and that the amplitudes of
individual pulsational signals also changed over time.

\begin{figure*}
\includegraphics[width=180mm,viewport=0 0 540 390]{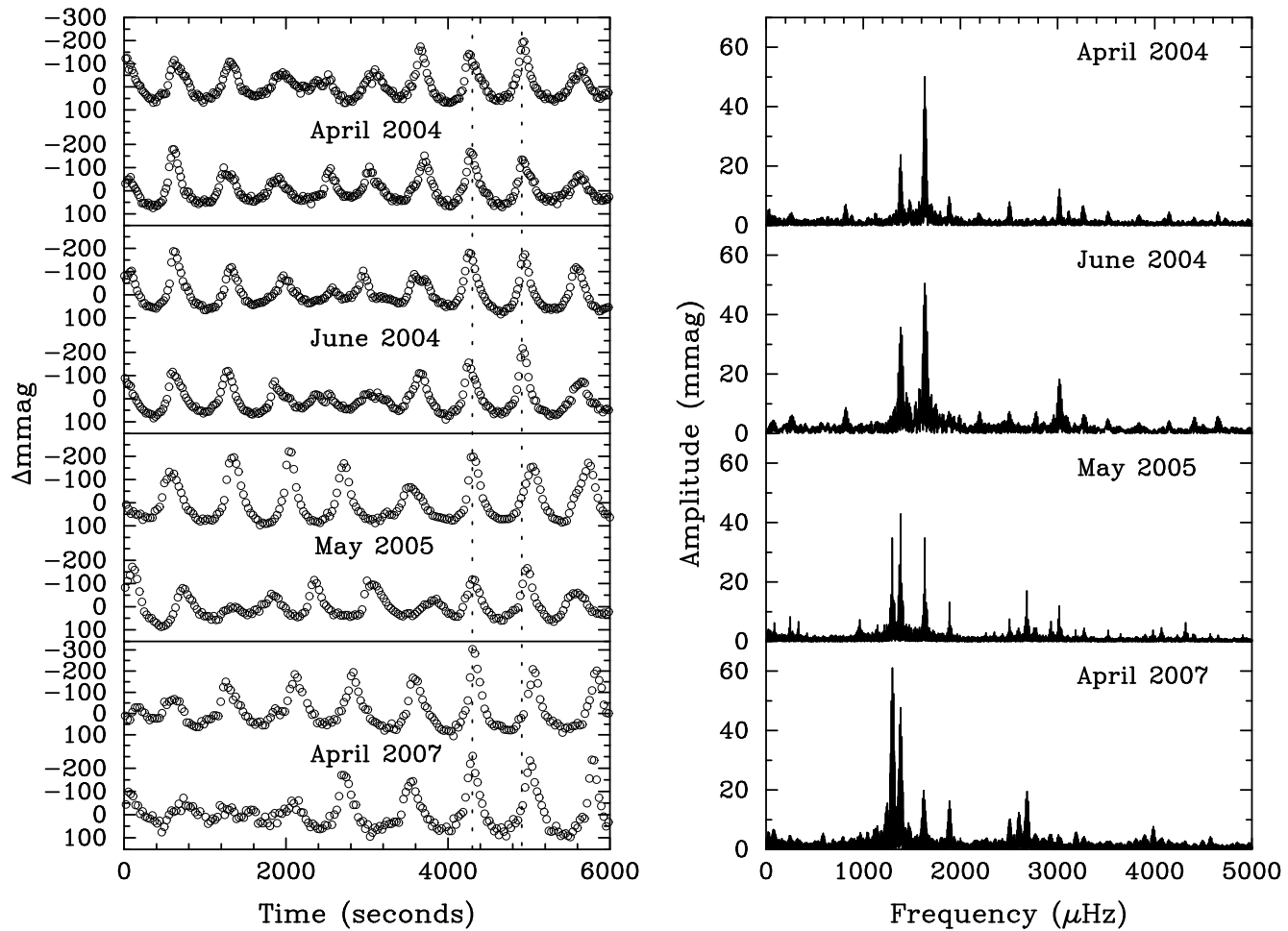}
\caption[]{Some example light curves (left-hand side) and the amplitude 
spectra of our data sets from the four different epochs. The dotted 
vertical lines denote the average distance between consecutive light 
maxima in the April 2004 data; note the temporal increase of the ``mean''
variability period directly visible in the light curves.}
\end{figure*}

Under these circumstances (individual data sets with time bases that are
short compared to their separations, often complicated window functions,
variable periods and amplitudes) it becomes clear that the frequency
analysis will be difficult and that the data sets cannot be analysed
jointly. We therefore used a different strategy: initiating the analysis
with the apparently best data subset, and then using the experience gained
for examining the other subsets.

\subsection{The May 2005 multisite data}

We started the analysis by computing Fourier amplitude spectra of the
light curves, and then determined the frequencies of the strongest
signals. These frequencies served as input parameters for a fit to the
light curves that was optimized by nonlinear least squares sine-wave
fitting implemented in {\tt Period98}. In case of ambiguities due to
aliasing, the frequency that yielded the lowest residuals between light
curve and fit after optimization was chosen as input. As the next step, we
prewhitened the fit from the light curve, computed the Fourier amplitude
spectrum of the residuals and used it to search for further periodic
signals.

After a few prewhitening steps it became clear that amplitude and/or
frequency variations had occurred during the multisite campaign. This
imposes severe complications on frequency analyses with Fourier
algorithms, whose fundamental assumption is that the amplitudes and
frequencies of the individual signals present are constant within the data
set. We therefore decided to use only the central part of the data set for
the purpose of frequency search. This core part of the run contained 69\%
of the campaign data, spanned 5.6~d, corresponding to a frequency
resolution of 2.1~$\mu$Hz and had a duty cycle of 53\%.

Using only the central part of the run, we carried on with the
prewhitening process as described above. Each signal detected during this
procedure was checked whether or not it was a combination frequency, i.e.\
whether it corresponded to a harmonic, sum, or difference frequency of
signals detected previously, within the frequency resolution of the data
set (Table 1). If so, its frequency was fixed to the exact sum or
difference of its parent signals within {\tt Period98}. This procedure is
valid here because the changes in the combination frequencies traced the
changes in the parent modes within the accuracy of our data.

In some cases, particularly for possible combinations of higher order,
ambiguities arose because more than one set of parent frequencies could be
matched to a given combination within the resolution of the data set. We
then chose these parent frequencies whose product of amplitudes was
highest, and required that the parent frequencies must have higher
amplitudes than the combination signal they generate.

Under these assumptions it is possible that the frequencies of 
some independent modes would be misidentified as combinations. However, we 
believe that disregarding a possible mode frequency is the safer choice 
compared to possibly accept some signals as real modes that are not.

We continued the analysis until no significant peak was left in the
residual amplitude spectrum, adopting the signal-to-noise criterion by
Breger et al.\ (1993, 1999). In brief, an independent signal must exceed an
amplitude signal-to-noise of 4 to be accepted, whereas $S/N>3.5$ is
sufficient for combination signals. We refer to the two papers mentioned
before for detailed discussions. In the end, 15 independent frequencies 
and 90 combination signals up to fifth order were detected.

\subsection{The April 2004 data}

Because of the long nightly time series acquired during our first
observing run on the star (more than nine hours per night on average),
this single-site data set is also well suited for frequency analyses. With
the same procedures as described before, we detected 12 independent and 44
combination frequencies in this data set, with no evidence for amplitude
and/or frequency variations throughout the observations.

\subsection{The June 2004 data}

Although acquired with the largest telescope we had available for this
study, this data set is not ideal for frequency analysis. The spectral
window function is complicated and the frequency resolution is poor (see
Fig.\ 1). Only the strongest signals could be detected without ambiguity,
and we had to resort to adopting frequencies determined from the April
2004 data to avoid aliasing problems when trying to reveal signals with
poorer signal-to-noise. Our frequency solution for this data set
containing 10 independent and 27 combination frequencies is probably
incomplete. On the other hand, we believe that we have resisted the
temptation to over-interpret this data set.

\subsection{The April 2007 data}

These measurements have the best formal frequency resolution, but due to
their sampling the spectral window is complicated. Still, we could
determine a large number of the signals in the light curves independently,
and only in a few cases we had to adopt frequencies known from analyses of
the other data sets. Amplitude and/or frequency variations likely
occurred during these observations as well, but not on a scale severely
hampering our analysis. Finally, we arrived at 11 independent and 40
combination frequencies for this data set.

\subsection{Putting it all together}

After the analyses of the individual data sets were completed, the results
were merged and the frequency solutions were compared. We list all 19
independent frequencies in our light curves in Table 2, all 52 first-order
combination frequencies in Table 3, all 60 third-order combination
frequencies in Table 4, and all 29 fourth-order combination frequencies
together with the 7 fifth-order combinations in Table 5.

\begin{table*}
\caption[]{Independent frequencies in our light curves of
EC\,14012$-$1446. The formal errors in the amplitudes are $\pm 0.4$~mmag
for April 2004, $\pm 0.5$~mmag for June 2004, $\pm 0.3$~mmag for May 2005
and $\pm 0.5$~mmag for April 2007}
\smallskip
\begin{tabular}{lcccccccc}
\hline
 & \multicolumn{2}{c}{April 2004} & \multicolumn{2}{c}{June 2004} & \multicolumn{2}{c}{May 2005} &\multicolumn{2}{c}{April 2007} \\
\hline
ID & Freq. & Ampl. & Freq. & Ampl. & Freq. & Ampl. & Freq. & Ampl. \\
 & ($\mu$Hz) & (mmag) & ($\mu$Hz) & (mmag) & ($\mu$Hz) & (mmag) & ($\mu$Hz) & (mmag)\\
\hline
$f_{1}$ &  821.26 $\pm$ 0.06 & 7.1 &  821.52 $\pm$ 0.13 & 7.8 \\
$f_{2}$ & 1132.89 $\pm$ 0.14 & 2.9 & \\
$f_{3}$ & & & & & 1289.28 $\pm$ 0.06 &   5.3 \\
$f_{4}$ & & & & & 1295.73 $\pm$ 0.04 &   8.7 \\
$f_{5}$ & & & & & 1299.00 $\pm$ 0.01 &  39.8 &  1300.62 $\pm$ 0.01 &  63.6 \\
$f_{6}$ & & & & & 1371.89 $\pm$ 0.04 &   9.1 &  1370.89 $\pm$ 0.01 &  20.0 \\
$f_{7}$ & 1375.53 $\pm$ 0.08 & 5.5  &  1375.02 $\pm$ 0.15 & 6.9 & 1375.60 $\pm$ 0.07 & 4.7\\
$f_{8}$ & & & & & 1381.97 $\pm$ 0.03 &   9.6 \\ 
$f_{9}$ & 1385.50 $\pm$ 0.02 & 24.4 & 1385.62 $\pm$ 0.03 &  35.1 & 1384.84 $\pm$ 0.01 &  40.6 & 1385.32 $\pm$ 0.01 &  44.1 \\
$f_{10}$ & & & & & 1394.91 $\pm$ 0.04 &   7.3 \\
$f_{11}$ & & & & & 1398.29 $\pm$ 0.03 &  10.9 &  1399.84 $\pm$ 0.02 &  13.6 \\
$f_{12}$ & 1464.17 $\pm$ 0.13 & 3.2  &  1464.17 $\pm$ 0.15 & 6.8 & & & 1463.95 $\pm$ 0.04 &   6.4 \\
$f_{13}$ & 1474.12 $\pm$ 0.04 & 9.3  &  1474.95 $\pm$ 0.11 & 9.1 & 1473.02 $\pm$ 0.06 &   5.6 & 1473.04 $\pm$ 0.04 &   6.3 \\
$f_{14}$ & 1484.13 $\pm$ 0.16 & 2.6 & \\
$f_{15}$ & 1623.28 $\pm$ 0.04 & 11.2 & 1623.51 $\pm$ 0.11 & 9.0 &   1623.86 $\pm$ 0.03 &  10.5 & 1623.20 $\pm$ 0.01 &  18.6 \\
$f_{16}$ & 1633.36 $\pm$ 0.01 & 48.1 & 1633.60 $\pm$ 0.02 &  48.3 & 1633.71 $\pm$ 0.01 &  33.5 & 1633.69 $\pm$ 0.05 &   4.7 \\
$f_{17}$ & 1643.40 $\pm$ 0.03 & 14.2 & 1642.96 $\pm$ 0.07 &  13.6 & 1643.03 $\pm$ 0.05 &   7.0 & 1644.08 $\pm$ 0.06 &   4.1 \\
$f_{18}$ & 1887.47 $\pm$ 0.05 & 8.9  &  1887.79 $\pm$ 0.11 & 9.3 & 1887.34 $\pm$ 0.03 &  12.3 & 1887.59 $\pm$ 0.02 &  15.2 \\
$f_{19}$ & 2504.86 $\pm$ 0.05 & 8.7  &  2504.98 $\pm$ 0.12 & 8.1 & 2504.97 $\pm$ 0.05 &   6.8 & 2504.65 $\pm$ 0.03 &   8.6 \\
\hline
\end{tabular}
\end{table*}

\begin{table*}
\caption[]{Second-order combination frequencies in our light curves of EC\,14012$-$1446}
\smallskip
\begin{tabular}{lcccccccccc}
\hline
 & \multicolumn{2}{c}{April 2004} & \multicolumn{2}{c}{June 2004} & \multicolumn{2}{c}{May 2005} &\multicolumn{2}{c}{April 2007} \\
\hline
ID & Freq. & Ampl. & Freq. & Ampl. & Freq. & Ampl. & Freq. & Ampl. \\
 & ($\mu$Hz) & (mmag) & ($\mu$Hz) & (mmag) & ($\mu$Hz) & (mmag)& ($\mu$Hz) & (mmag)\\
\hline
$f_{6}-f_{5}$ & & & & & & & 70.27 & 3.9 \\
$f_{9}-f_{5}$ & & & & &  85.83 & 5.9 & 84.69 & 6.7 \\
$f_{12}-f_{5}$ & & & & & & &  163.33 & 3.1 \\
$f_{16}-f_{9}$  &    247.86 &   2.8   &    247.98 &   4.9 &  248.87 & 7.1 &  \\
$f_{15}-f_{6}$ & & & & & & & 252.31 & 3.9 \\
$f_{18}-f_{16}$ &    254.12 &   3.6  & \\
$f_{17}-f_{7}$ & & &    267.94 &   5.9 & \\
$f_{15}-f_{5}$ & & & & &  & &  322.58 & 2.9 \\
$f_{16}-f_{5}$ & & & & &  334.71 & 6.0 &  \\
$f_{17}-f_{5}$ & & & & & & &  343.45 & 3.0 \\
$f_{18}-f_{9}$ & & & & &  502.50 & 1.8 &  502.27 & 2.5 \\
$f_{18}-f_{5}$ & & & & &  588.34 & 1.9 &  586.97 & 5.0 \\
$f_{19}-f_{15}$ & & & & & & &  881.44 & 3.5 \\
$f_{19}-f_{6}$ & & & & & & & 1133.75 & 5.2 \\
$f_{1}+f_{7}$  &   2196.79 &   3.3   &   2196.55 &   5.4 & \\
$f_{1}+f_{9}$  &   2206.76 &   2.4  & \\
$f_{1}+f_{13}$  &   2295.39 &   1.6   &   2296.48 &   2.7 & \\
$f_{1}+f_{16}$  &   2454.62 &   3.0   &   2455.13 &   4.1 & \\
$2f_{5}$ & & & & & 2598.00 & 4.6 &  2601.25 &  12.8 \\
$f_{5}+f_{6}$ & & & & & & &  2671.52 & 7.5 \\
$f_{5}+f_{9}$ & & & & & 2683.84 &  17.3 &  2685.94 &  18.8 \\
$f_{5}+f_{10}$ & & & & & 2693.91 & 3.1 &  \\
$f_{5}+f_{11}$ & & & & & 2697.29 & 2.9 & 2700.47 & 5.6 \\
$f_{6}+f_{9}$ & & & & & 2756.72 & 3.7 & 2756.21 & 4.6 \\
$f_{5}+f_{12}$ & & & & & & & 2764.58 & 3.8 \\
$2f_{9}$       &   2770.99 &   1.7   &   2771.25 &   5.8 & 2769.67 & 2.3 & 2770.64 & 3.7 \\
$f_{5}+f_{13}$ & & & & & 2772.02 & 3.0 &  \\
$f_{9}+f_{11}$ & & & & & 2783.12 & 1.6 & 2785.16 & 3.0 \\
$f_{6}+f_{13}$ & & & & & 2844.91 & 2.0 &  \\
$f_{9}+f_{13}$  &   2859.62 &   2.8   &   2860.58 &   2.1 & \\
$f_{5}+f_{15}$ & & & & & & & 2923.83 & 4.7 \\
$f_{4}+f_{16}$ & & & & & 2929.43 & 2.0 &  \\
$f_{5}+f_{16}$ & & & & & 2932.71 & 8.5 &  \\
$f_{6}+f_{16}$ & & & & & 3005.60 & 3.7 &  \\
$f_{9}+f_{15}$ & & & & & 3008.70 & 1.8 & 3008.52 & 4.8 \\
$f_{9}+f_{16}$  &   3018.85 &  12.3   &   3019.23 &  15.7 & 3018.54 &  10.4 &  \\
$f_{11}+f_{15}$ & & & & & 3022.15 & 3.3 & 3023.05 & 3.7 \\
$f_{9}+f_{17}$ &   3028.89 &   3.5   &   3028.58 &   5.4 & 3027.86 & 2.0 &  \\
$f_{13}+f_{16}$  &   3107.48 &   5.0   &   3108.56 &   4.1 & \\
$f_{13}+f_{17}$ &   3117.52 &   1.6  & \\
$f_{5}+f_{18}$ & & & & & 3186.34 & 3.4 & 3188.22 & 5.0 \\
$f_{15}+f_{16}$  &   3256.63 &   3.1   &   3257.12 &   3.5 & \\
$2f_{16}$       &   3266.71 &   6.2   &   3267.21 &   6.2 & 3267.42 & 2.8 &  \\
$f_{9}+f_{18}$  &   3272.97 &   1.9 & & & 3272.18 & 3.2 & 3272.91 & 4.4 \\
$f_{16}+f_{17}$  &   3276.75 &   4.0   &   3276.56 &   3.7 & \\
$f_{16}+f_{18}$  &   3520.83 &   4.2   &   3521.39 &   4.6 & 3521.05 & 4.2 &  \\
$f_{5}+f_{19}$ & & & & &  & & 3805.27 & 2.8 \\
$f_{9}+f_{19}$  &   3890.36 &   1.7  & \\
$f_{11}+f_{19}$ & & & & & 3903.26 & 2.1 & 3904.49 & 3.1 \\
$f_{16}+f_{19}$  &   4138.22 &   2.0   &   4138.58 &   2.8 & \\
$f_{17}+f_{19}$ &   4148.26 &   4.0   &   4147.94 &   3.4 & \\
$f_{18}+f_{19}$ & & & & & 4392.31 & 1.0 & \\
\hline
\end{tabular}
\end{table*}

\begin{table*}
\caption[]{Third-order combination frequencies in our light curves of EC\,14012$-$1446}
\smallskip
\begin{tabular}{lcccccccc}
\hline
 & \multicolumn{2}{c}{April 2004} & \multicolumn{2}{c}{June 2004} & \multicolumn{2}{c}{May 2005} &\multicolumn{2}{c}{April 2007} \\
\hline
ID & Freq. & Ampl. & Freq. & Ampl. & Freq. & Ampl. & Freq. & Ampl. \\
 & ($\mu$Hz) & (mmag) & ($\mu$Hz) & (mmag) & ($\mu$Hz) & (mmag) & ($\mu$Hz) & (mmag)\\
\hline
$2f_{5}-f_{17}$ & & & & & & & 957.17 & 3.4 \\
$2f_{4}-f_{16}$ & & & & &     957.74 &   2.2 &   \\
$f_{4}+f_{5}-f_{16}$ & & & & & 961.02 &   4.3 &   \\
$2f_{5}-f_{16}$ & & & & &  964.30 &   8.9 &   \\
$2f_{5}-f_{15}$ & & & & &   974.14 &   3.8 &   \\
$f_{5}+f_{9}-f_{16}$ & & & & &  1050.13 &   3.3 &   \\
$f_{5}+f_{9}-f_{15}$ & & & & &  1059.98 &   2.5 &   \\
$2f_{9}-f_{17}$ & & & & &   1126.64 &   2.7 &   \\
$2f_{9}-f_{16}$ & & & & &   1135.96 &   3.5 &   \\
$2f_{5}-f_{12}$ & & & & & & & 1137.30 & 4.8 \\
$2f_{9}-f_{15}$ & & & & &   1145.81 &   3.2 &   \\
$f_{5}+f_{6}-f_{12}$ & & & & & & & 1207.57 & 4.4 \\
$f_{4}+f_{5}-f_{9}$ & & & & &  1209.89 &   2.0 &   \\
$2f_{5}-f_{9}$ & & & & &   1213.17 &   2.3 & 1215.93 & 4.9 \\
$2f_{5}-f_{6}$ & & & & & & & 1230.36 & 8.9 \\
$f_{5}+f_{9}-f_{11}$ & & & & &  1285.55 &   6.6 &   \\
$f_{5}+f_{10}-f_{9}$ & & & & &  1309.07 &   6.1 &   \\
$f_{5}+f_{9}-f_{6}$ & & & & &  & & 1315.05 & 4.5 \\
$2f_{16}-f_{18}$ &   1379.24 &   3.1   & \\
$2f_{9}-f_{5}$ & & & & &  1470.67 &   3.6 &   \\
$f_{9}+f_{16}-f_{13}$ & & & & &  1545.53 &   2.1 &   \\
$f_{5}+f_{16}-f_{9}$ & & & & &  1547.88 &   2.7 &   \\
$f_{9}+f_{17}-f_{13}$ & & & & &   1554.84 &   2.3 &   \\
$f_{11}+f_{16}-f_{9}$ & & & & &   1647.16 &   6.0 &   \\
$f_{9}+f_{16}-f_{5}$ & & & & &  1719.54 &   1.7 &   \\
$2f_{16}-f_{13}$  &   1792.59 &   2.3   &   1792.26 &   2.4  & \\
$f_{15}+f_{16}-f_{9}$  &   1871.14 &   2.1   & \\
$2f_{16}-f_{9}$ &   1881.22 &  5.0  & 1881.58 &  7.4  & 1882.58 &  1.8 & \\
$f_{16}+f_{17}-f_{9}$ & & & & &  1891.90 &   3.2 &   \\
$2f_{16}-f_{6}$ & & & & & 1895.53 &   2.0 &   \\
$2f_{16}-f_{5}$ & & & & & 1968.42 &   0.8 &   \\
$f_{16}+f_{19}-f_{9}$ & & & & &  2753.84 &   2.3 &   \\
$f_{1}+f_{13}+f_{16}$ &   3830.15 &   3.2   &   3830.15 &   3.0  & \\
$f_{1}+f_{9}+f_{17}$ &   3850.16 &   1.0   & \\
$3f_{5}$ & & & & & & & 3901.87 & 3.8 \\
$2f_{5}+f_{9}$ & & & & &  3982.84 &   3.7 & 3986.57 & 7.7 \\
$f_{5}+2f_{9}$ & & & & &  4068.67 &   3.9 & 4071.26 & 3.3 \\
$2f_{5}+f_{13}$ & & & & & & & 4074.29 & 3.9 \\
$f_{5}+f_{9}+f_{11}$ & & & & & 4082.12 &   1.6 &   \\
$f_{1}+2f_{16}$       &   4087.98 &   2.2   & \\
$2f_{5}+f_{16}$ & & & & & 4231.71 &   2.8 &   \\
$f_{5}+f_{9}+f_{16}$ & & & & & 4317.55 &   6.7 &   \\
$f_{5}+f_{10}+f_{16}$ & & & & & 4327.62 &   1.5 &   \\
$2f_{9}+f_{16}$ &   4404.35 &   2.2   &   4404.85 &   5.3  & 4403.38 &   1.6 & \\
$f_{5}+f_{13}+f_{16}$ & & & & & 4405.73 &   1.5 &   \\
$f_{9}+f_{12}+f_{16}$  &   4483.02 &   0.8   & \\
$2f_{5}+f_{18}$ & & & & & & & 4488.84 & 2.5 \\
$f_{9}+f_{13}+f_{16}$  &   4492.98 &   1.6   &   4494.18 &   2.3  & \\
$f_{5}+2f_{16}$ & & & & &  4566.42 &   1.6 &   \\
$f_{5}+f_{9}+f_{18}$ & & & & & 4571.18 &  2.2 & 4573.54 & 3.7 \\
$f_{9}+2f_{16}$ &   4652.21 &   4.9   &   4652.83 &   5.1  & 4652.25 &   2.0 & \\
$f_{9}+f_{16}+f_{17}$ &   4662.25 &   1.5   &   4662.19 &   2.1  & \\
$f_{1}+f_{9}+f_{19}$ &   4711.62 &   0.9   & \\
$f_{13}+2f_{16}$       &   4740.83 &   1.7   & \\
$f_{5}+f_{16}+f_{18}$ & & & & & 4820.05 &   1.5 &   \\
$3f_{16}$ &   4900.07 &   1.9   & \\
$f_{9}+f_{16}+f_{18}$ & & & & &  4905.88 &   2.0 &   \\
$f_{13}+f_{16}+f_{17}$       &   5473.54 &   2.6   & \\
$f_{9}+f_{16}+f_{19}$ & & &   5524.21 &   1.8  & \\
$f_{16}+f_{17}+f_{19}$       &   5781.61 &   1.6   & \\
\hline
\end{tabular}
\end{table*}

\begin{table*}
\caption[]{Fourth and fifth-order combination frequencies in our light curves of EC\,14012$-$1446}
\smallskip
\begin{tabular}{lcccccccc}
\hline
 & \multicolumn{2}{c}{April 2004} & \multicolumn{2}{c}{June 2004} & \multicolumn{2}{c}{May 2005} &\multicolumn{2}{c}{April 2007} \\
\hline
ID & Freq. & Ampl. & Freq. & Ampl. & Freq. & Ampl. & Freq. & Ampl. \\
 & ($\mu$Hz) & (mmag) & ($\mu$Hz) & (mmag) & ($\mu$Hz) & (mmag) & ($\mu$Hz) & (mmag)\\
\hline
$f_{9}+f_{16}-2f_{13}$ & & &     69.32 &   5.5 & \\
$2f_{9}-f_{5}-f_{11}$ & & & & &  72.38 &   1.8 & \\
$f_{9}+f_{16}-2f_{5}$ & & & & & 420.54 &   2.6 & \\
$2f_{16}-2f_{5}$ & & & & &   669.41 &   1.8 & \\
$2f_{16}-f_{4}-f_{5}$ & & & & &  672.69 &   2.1 & \\
$f_{1}+f_{9}+f_{17}-f_{16}$  &   2216.80 &   2.1  & \\
$3f_{5}-f_{16}$ & & & & & 2263.30 &   2.7 & \\
$f_{4}+f_{5}+f_{9}-f_{16}$ & & & & & 2345.85 &   2.2 & \\
$2f_{5}+f_{9}-f_{16}$ & & & & &  2349.13 &   3.3 & \\
$f_{5}+f_{8}+f_{9}-f_{16}$ & & & & &  2432.10 &   2.1 & \\
$f_{5}+2f_{9}-f_{16}$ & & & & &  2434.97 &   2.0 & \\
$2f_{5}+f_{9}-f_{6}$ & & & & & & & 2615.68 & 3.2 \\
$2f_{5}+f_{15}-f_{6}$ & & & & & & & 2853.56 & 3.1 \\
$2f_{9}+f_{13}-f_{5}$ & & & & &   2943.69 &   2.2 & \\
$f_{4}+2f_{16}-f_{9}$ & & & & &   3178.31 &   1.2 & \\
$3f_{16}-f_{9}$              &   3514.57 &   2.7   &   3515.19 &   4.3 & \\
$f_{1}+2f_{9}+f_{16}$        &   5225.61 &   1.7  & \\
$3f_{5}+f_{9}$ & & & & &  & & 5287.19 & 2.7 \\
$2f_{5}+2f_{9}$ & & & & &   5367.68 &   1.8 & \\
$2f_{5}+f_{9}+f_{16}$ & & & & &  5616.55 &   2.5 & \\
$f_{5}+2f_{9}+f_{16}$ & & & & &  5702.38 &   1.9 & \\
$3f_{9}+f_{16}$              &   5789.85 &   1.2  & \\
$f_{5}+f_{9}+2f_{16}$ & & & & &  5951.26 &   1.0 & \\
$f_{5}+2f_{9}+f_{18}$ & & & & &  5956.01 &   1.0 & \\
$2f_{9}+2f_{16}$ & & &   6038.46 &   3.0 & \\
$f_{5}+f_{9}+f_{16}+f_{18}$ & & & & &  6204.89 &   1.0 & \\
$3f_{16}+f_{9}$              &   6285.57 &   2.0  & \\
$f_{9}+2f_{13}+f_{16}$  &   6689.76 &   1.1  & \\
$f_{9}+f_{13}+f_{16}+f_{17}$ &   6859.04 &   1.4  & \\
\hline
$2f_{5}+f_{9}-f_{11}-f_{16}$ & & & & &    950.85 &   2.6 \\
$3f_{9}-f_{5}-f_{18}$ & & & & &    968.17 &   1.8 \\
$3f_{5}+f_{9}-f_{16}$ & & & & &   3648.13 &   2.3 \\
$f_{5}+f_{11}+f_{16}+f_{19}-f_{18}$ & & & & &   4948.63 &   1.3 \\
$3f_{5}+2f_{9}$ & & & & &   6666.68 &   1.1 \\
$3f_{5}+f_{9}+f_{16}$ & & & & &   6915.55 &   0.8 \\
$2f_{5}+2f_{9}+f_{16}$ & & & & &   7001.39 &   1.1 \\
\hline
\end{tabular}
\end{table*}

\section{Discussion}

\subsection{The independent frequencies}

As can be seen in Table 2, some, but not all of the independent
frequencies are consistently present in all data sets. Some are singlets,
and some show (regular) frequency splittings. This requires individual
discussions of all these signals or signal groups. Again, we start with
the simplest cases and move on to the more difficult ones, choosing to
proceed from the highest frequencies to the lowest.

The signal at 2504~$\mu$Hz is well-behaved. It is present throughout all
data sets, with its frequency and amplitude roughly staying constant, and
it shows no splittings. It is the best candidate to search for possible
evolutionary period changes, but our data are not sufficient to reach a
definite conclusion in this respect. We note, and caution, that in the
discovery data by Stobie et al.\ (1995) this signal had an amplitude
60\% higher than in our measurements.

The 1887~$\mu$Hz variation is reasonably stable as well. However, in most
data sets it had a companion near 1881~$\mu$Hz, which is identified with a
third-order combination frequency. The amplitude of the 1887~$\mu$Hz
signal increased with time. This finding must be taken with caution
because there is no information about the star's behaviour between the
different data sets. This signal also showed a higher amplitude in the
Stobie et al.\ (1995) data than in any of our data sets.

A frequency triplet centred at 1633~$\mu$Hz is present in all data sets.  
This triplet is always equally spaced within the errors, but the amount of
the splitting varied: from 10.06~$\mu$Hz in April 2004 to
9.73~$\mu$Hz in June 2004, to 9.59~$\mu$Hz in May 2005 and to
10.44~$\mu$Hz in April 2007. Given the amount by which this splitting 
changes and the observational errors, this variation is probably intrinsic 
to the star.

An independent frequency at 1474~$\mu$Hz is also detected in each
individual data set. In May 2005 it was a singlet, in June 2004 and April
2007 it was accompanied by another independent frequency lower by 10.8 and
9.1~$\mu$Hz, respectively, and in April 2004 it even was the centre of a
triplet split by 9.98~$\mu$Hz. Within the errors, this splitting is
consistent with the separation of the triplet around 1633~$\mu$Hz in the
same period of observation (April 2004).

A strong signal at 1385~$\mu$Hz is present in all data sets. In April and
June 2004 it had a companion at a frequency 9.97 and 10.60~$\mu$Hz lower,
respectively. However, in April 2007 the 1385~$\mu$Hz signal was the
centre of an equally spaced frequency triplet separated by 14.48~$\mu$Hz.
In May 2005, the situation is even more difficult: we found complicated
structure in this frequency region, with altogether six peaks and with the
presence of two triplets centred on 1385~$\mu$Hz with both the $\approx
10$ and $\approx 14$~$\mu$Hz splittings. We note that the separation of
some of the neighbouring signals is close to the resolution of our data
set (1.5/$\Delta T$, where $T$ is the time base of the data, Loumos \&
Deeming 1978), but these signals remain at the same frequencies when the
full campaign data set is considered.

At the two later observing epochs, a strong signal at 1300~$\mu$Hz
emerged. It was the strongest peak the amplitude spectrum in April 2007 as
a singlet, and in May 2005 it had two lower-frequency companions separated
by 3.3 and 9.7~$\mu$Hz, respectively. There was no trace of it in our
measurements from 2004, and it was not detected by Stobie et al.\ (1995).

The two lowest-frequency independent variations in our light curves were
only present during (part of) 2004: a singlet at 1133~$\mu$Hz in April,
and another singlet at 821~$\mu$Hz in both observing periods in 2004. We 
note that although numerically consistent in June 2004, an interpretation 
of $f_1$ as the subharmonic of $f_{17}$ generated a poor fit to the data 
and was therefore rejected.

Stobie et al.\ (1995) reported an additional independent frequency of
1067~$\mu$Hz in their discovery data. However, it was only present in
their first night of observation, and we cannot confirm its presence in
any of our data sets. As we do not have the data by Stobie et al.\ (1995)
at our disposal, we will not discuss this signal any further.

\subsubsection{Short-term amplitude and frequency variations}

As noted before, we had to concentrate our frequency analysis of the May
2005 multisite photometry on the central part with the highest duty cycle
because amplitude variations had occurred during the observations. After
prewhitening the multifrequency solution listed in Tables $2 - 5$ from
these data, we still find residual mounds of amplitude left around most of 
the independent frequencies, most notably in the region with the most 
complicated structure around 1385~$\mu$Hz.

This raises the suspicion that at least some of the signals in this
frequency domain are artefacts due to amplitude and/or frequency
variability. However, tests to check this hypothesis remained
inconclusive. Consequently, it is not clear whether or not all the signals
around 1385~$\mu$Hz (Table 2) indeed correspond to normal mode
frequencies.

We examined the occurrence of amplitude and frequency variations in the
May 2005 data. We used the full data set for this purpose, and selected
six independent signals, either singlets or the strongest multiplet
components. First, all but these signals were prewhitened from the light
curves, using Tables $2 - 5$ as input and assuming constant amplitude and
phase for all variations. We subdivided the residual light curves into
sections long enough to avoid beating within the six independent signals
under consideration and computed the amplitudes and phases for each.
These are represented in graphical form in Fig.\ 3.

\begin{figure*}
\includegraphics[width=180mm,viewport=00 00 520 445]{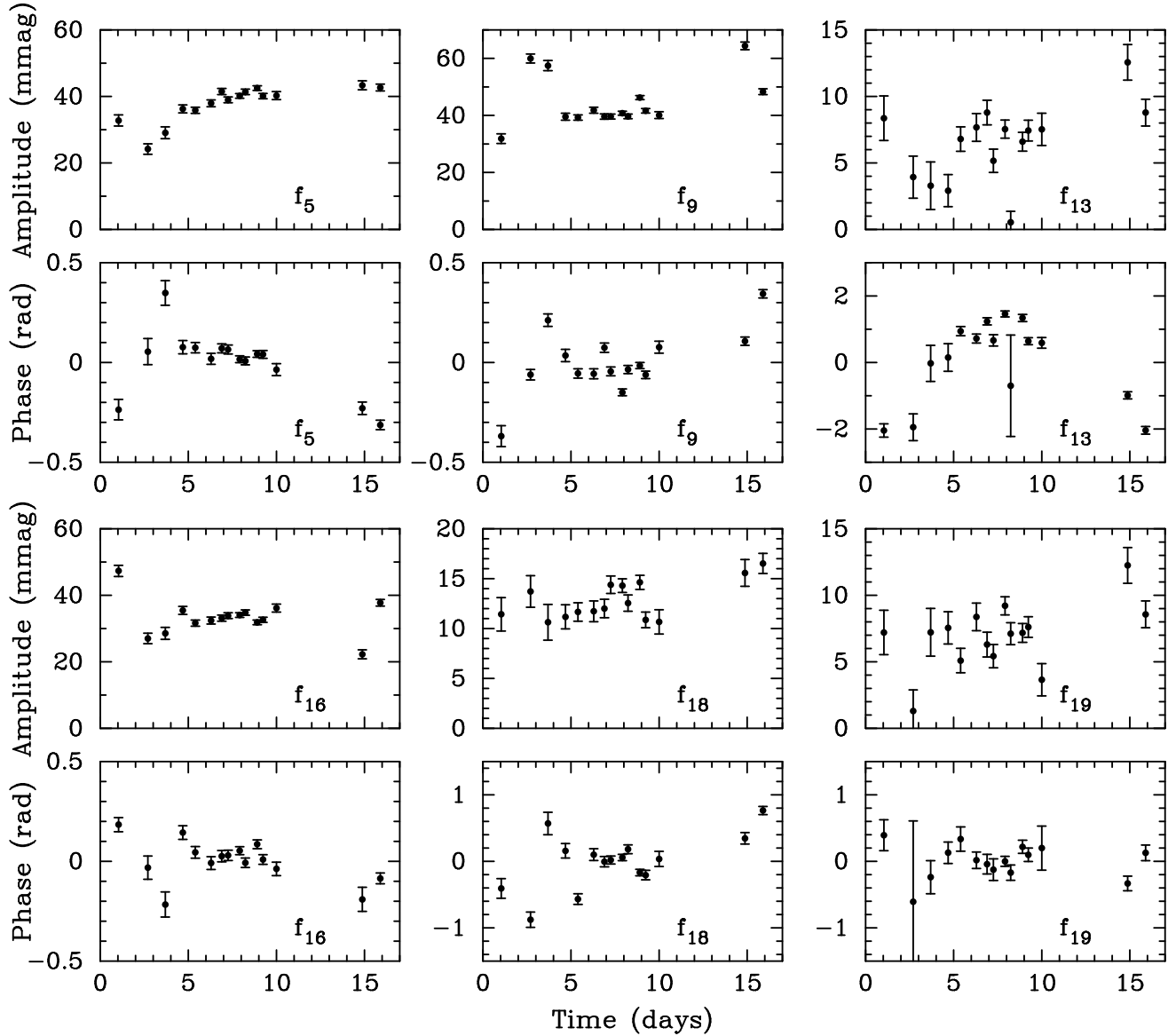}
\caption[]{Amplitude and frequency variations of six independent signals 
in the light curve of EC~14012$-$1446 during the observations in May 2005. 
The error bars of the individual data points are 1$\sigma$.
The panels are grouped with respect to the signal, and have been arranged 
to compare the amplitude and phase behaviour directly. Note the different 
ordinate scales in the different panels. The central subset of data used 
for frequency analysis corresponds to 4 $<$ Time $<$ 10 in this graph.}
\end{figure*}

It is hard to say something quantitatively about the amplitude and 
frequency changes (which would require an even higher observational duty
cycle), but we can make some qualitative statements.
\begin{itemize} 
\item All independent signals do show amplitude and frequency modulation, 
with the possible exception of $f_{19}$.
\item In general, signals at lower frequency (and therefore higher 
radial overtone) show more erratic behaviour. 
\item Amplitude variations are always accompanied by phase (frequency) 
variations. Whenever the amplitude is constant, the frequency is also 
constant.
\end{itemize}
Figure 3 also confirms that the amplitudes and phases of the variations 
remained constant during the central part of the observations, validating 
the most important assumption of the frequency analysis in Sect.\ 3.1. 

Finally, we point out the interesting behaviour of signal $f_{13}$ (top
right panels in Fig.\ 3): it showed two phase "jumps" of $\approx
\pi$~rad, at the beginning and the end of the observations. This might be
an effect of beating between closely spaced signals. The amplitude
spectrum of the full data set indeed shows two peaks of similar amplitude
spaced by 1.3~$\mu$Hz at this frequency. However, such a close frequency
doublet is not present at a significant level in the measurements from
2007 that would also have the time base to resolve it.

\subsection{The second-order combination frequencies}

The combination frequencies appearing in Fourier analysis of the light
curves of pulsating white dwarf stars are most commonly interpreted as
nonlinear distortions, most likely originating in their surface convection
zones (Brickhill 1992). However, an alternative interpretation is
resonantly coupled modes (Dziembowski 1982). These two hypotheses are
often difficult to separate, but as a general rule of thumb resonantly
excited modes should have considerably larger amplitudes compared to
signals that reflect light-curve distortions.

Figure 4 shows the ratio of the amplitudes of the individual combination
frequencies with respect to the product of the amplitudes of the parent
signals. Note that these amplitude ratios are given in units of
mag$^{-1}$, whereas the amplitudes themselves are quoted in
millimagnitudes. Whereas 85\% of the data points are located in an
interval of $8.5 \pm 6.2$ mag$^{-1}$, the remaining combinations show
considerably larger amplitude ratios.

\begin{figure*}
\includegraphics[width=180mm,viewport=00 00 570 199]{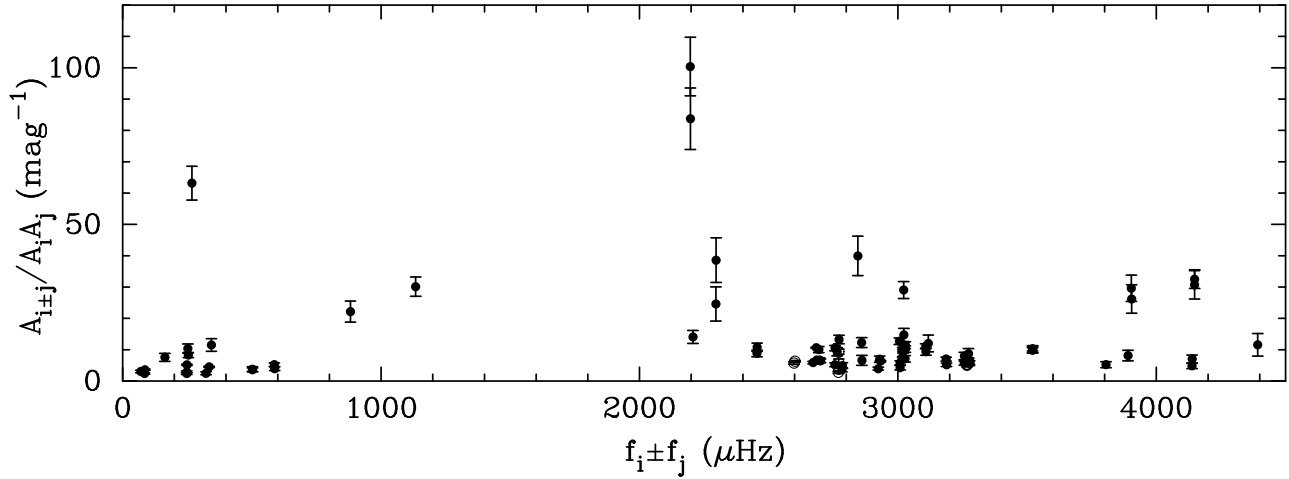}
\caption[]{Amplitude ratios between the second-order combination 
frequencies and the product of their parent signal amplitudes, with 
1$\sigma$ error bars indicated. Multiple determinations at a single 
frequency are due to combinations present in more than one data set.}
\end{figure*}

By far the largest amplitude ratio is due to the combination frequency at
2196~$\mu$Hz present in both data sets from the year 2004. It therefore is
our best candidate for a resonantly coupled mode. In addition, the
second-order combination at 2296~$\mu$Hz also stands out in Fig.\ 4. Both
of these combinations include the lowest-frequency independent signal
$f_1=821\,\mu$Hz. This latter signal also has two other combinations with
rather large amplitude ratio with respect to its parent variations (at
2207 and 2455\,$\mu$Hz, respectively). The third combination that shows
enhanced amplitude in both data sets from 2004 is the one at highest
frequency, involving $f_{19}$. Finally, the combination frequency
difference at 268~$\mu$Hz in June 2004 also has large amplitude, but the
residual amplitude spectrum in this data set has high noise around this
frequency. We therefore suspect that the large amplitude of the
268~$\mu$Hz signal may be spurious.

Turning to the data sets from 2005 and 2007, we do not find such extreme
amplitude ratios. Three of the five high-amplitude combination frequencies
again involve $f_{19}$, and the other two contain ``satellite''
frequencies around the strong signal at $f_9 \approx 1385~\mu$Hz. Of
highest interest is the combination at 1133.75~$\mu$Hz in the April 2007
data. Its frequency is consistent within the errors with that of the
independent signal $f_2$ that was only present in the year 2004. However,
in 2004 only one of the parents of the combination present in 2007 was
detected. Therefore $f_2$ must be regarded as an independent signal in the
data from 2004.

We remind that during the four epochs of observations of EC~14012$-$1446
at our disposal, amplitude variations occurred. This puts us into a
position to examine whether or not the amplitudes of combination signals
present in more than one of these data sets followed the corresponding
changes of their parent signals. We restrict this analysis to the
combination frequency sums because these are less affected by
observational noise, i.e.\ have better signal-to-noise on average.

In order not to make any assumptions about the origin of the combination
frequencies, we chose to compare the individual amplitude ratios of the
combinations that occur in more than one data set with respect to their
mean and their deviations from their formal accuracies. If there were no
changes in the relative amplitudes, we would expect a normal distribution
to result. The outcome of this test can be found in Fig.\ 5.

\begin{figure}
\includegraphics[width=88mm,viewport=00 00 365 230]{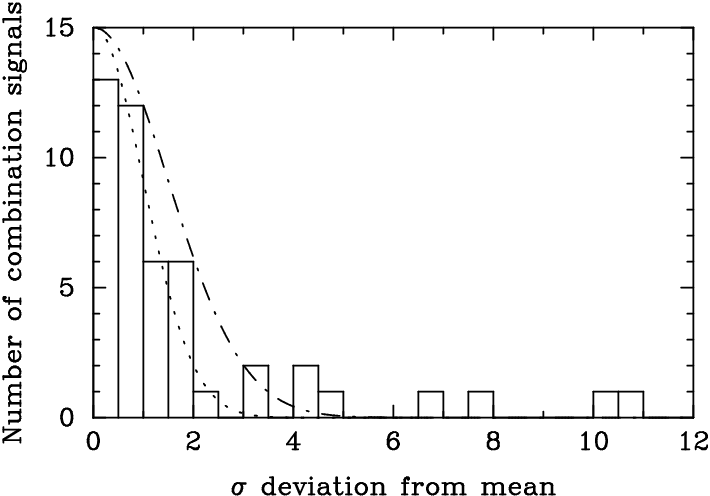}
\caption[]{The distribution of the deviation of the amplitude ratios of
recurring second-order combination frequencies with respect to their 
mean and formal errors (histogram bars). For comparison, two Gaussians 
with standard deviations of 1$\sigma$ (dotted line) and 1.5$\sigma$ 
(dash-dotted line) are shown.}
\end{figure}

Most of the amplitude ratios follow a Gaussian distribution with a width
of $\approx 1.2\sigma$. Given the fact that formal errors of the 
determinations of amplitudes tend to underestimate the real errors, this 
is not surprising.

There are nine recurring second-order combination signals whose relative
amplitude varies significantly, and they correspond to four different sets
of parents: $f_5+f_9$, $f_6+f_9$, $2f_9$, and $f_9+f_{16}$. All these
combinations contain the mode $f_9$, but we failed to find any systematics
in the temporal behaviour of the relative amplitudes of its combinations
due to the sparsity of data.

We also tested whether possible interactions with other combinations that
happen to have frequencies close to the four signals under investigation
might be able to affect their amplitudes. The result of this test was
negative because in general all possible alternative parent frequency
combinations have much lower amplitudes than the parents we assigned. The
single exception to this rule is $2f_9$ in the 2007 data set, whose
amplitude could be mildly affected by interference with $f_6+f_{11}$.

\subsection{The third-order combination frequencies}

The third-order combination frequencies determined can be separated into
two approximately equally populated groups: combinations of the form
$f_A+f_B+f_C$ and $f_A+f_B-f_C$, respectively. An important fact about the
second type of combination signals it that their frequencies are in the
same range as those of the independent signals (see also Vuille et al.\
2000). Consequently, they may be mistaken for independent frequencies if
not systematically searched for.

As a matter of fact, such confusion has occurred in the preliminary report
on EC~14012$-$1446 by Handler \& Romero-Colmenero (2005), where two
signals now identified as third-order combination frequencies were
believed to be, and interpreted as, independent mode frequencies. We
illustrate the severity of this problem in Fig.\ 6, that shows the
schematic amplitude spectrum of our multisite data set with the
independent frequencies and combination signals separated. Most of the
signals in this frequency range in fact correspond to combination
frequencies!

\begin{figure*}
\includegraphics[width=180mm,viewport=00 02 540 190]{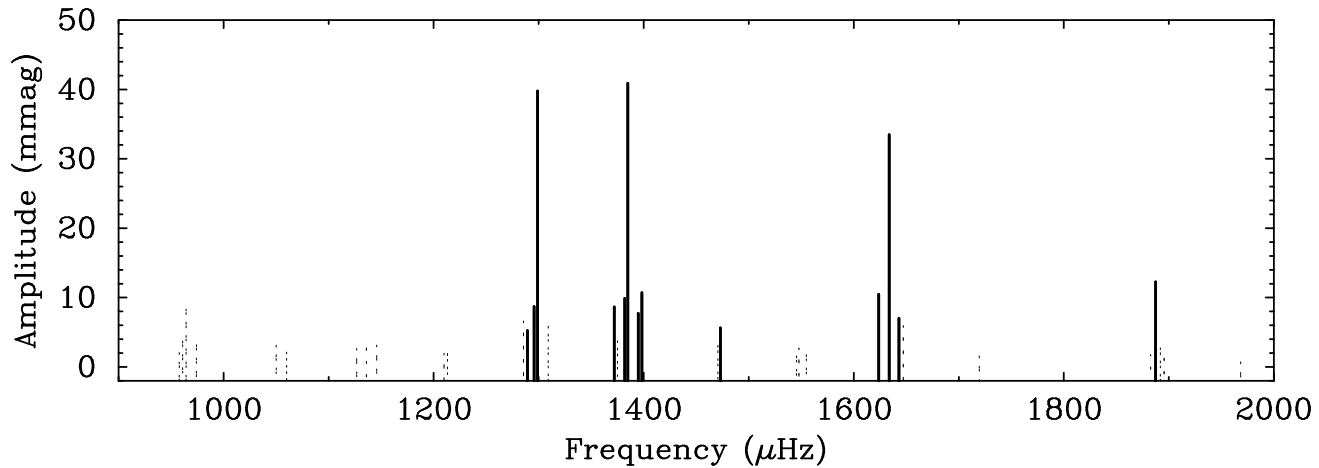}
\caption[]{Schematic amplitude spectrum of the May 2005 data (see Tables 
2 and 4). The independent frequencies are the thick full lines, whereas 
the third-order combination frequencies are shown with thin dotted lines.}
\end{figure*}

We again stress that we prefer to misinterpret a peak in the amplitude
spectrum as a combination signal as opposed to falsely taking combination
frequencies as real modes; we followed this guideline throughout the
frequency analysis. We will discuss the astrophysical implications of
confusing combination signals with independent mode frequencies below.

\subsection{The fourth and fifth-order combination frequencies}

We were able to identify several high-order combination frequencies in the
data. To our knowledge, this is the first report of the detection of
fifth-order combination signals in the light curves of a pulsating white
dwarf star. The identification of the signals that are pure frequency sums
should be secure because there are no ambiguities in assigning the
parents, whereas some identifications involving frequency sums and
differences may be incorrect. However, it is important to point out that
we only matched significant peaks in the amplitude spectrum, and that we
succeeded in explaining them all.

\section{Interpretation}

\subsection{The independent frequencies}

With our approach to frequency analysis, most of the independent
frequencies should correspond to normal mode oscillations. We interpret
$f_1$, $f_2$, $f_{18}$ and $f_{19}$ as single modes. The triplets $f_{12},
f_{13}, f_{14}$ and $f_{15}, f_{16}, f_{17}$ are normal mode multiplets;
their splitting of $\approx 10\, \mu$Hz is most likely due to rotation. We
believe that frequency $f_9$ is also caused a normal mode. It does show a
$\approx 10\, \mu$Hz splitting at times, but from time to time a $\approx
14.5\, \mu$Hz splitting is present. The signal $f_5$ is also suggested to
be caused by a normal mode, and it once showed a companion frequency again
separated by $\approx 10\, \mu$Hz.

Assuming that the modes where we found rotational splitting have a
spherical degree $\ell=1$, we can estimate the rotation rate of the star.  
The average amount of the splitting we believe to be of rotational origin
is $9.94\pm0.12\,\mu$Hz. Taking into account that this is about 4\% off
the asymptotic value (Brassard et al.\ 1992) and that the rotational
splitting is somewhat variable, we arrive at a rotation frequency of
$19.1\pm0.8\,\mu$Hz or at a fairly short rotation period of
$0.61\pm0.03$\,d.

Some of the independent signals in our light curves ($f_4, f_6, f_8,
f_{11}$) did not conform to rotationally split peaks as interpreted
before. There may be two possibilities for their origin. Firstly, they
could be members of rotationally split multiplets with different $\ell$.  
However, no reasonable assumption (either the $\approx 10\,\mu$Hz
splitting is due to $\ell=1$ modes or it is due to the $\ell=2$ rotational
splitting) in this framework can explain the whole set of signals detected
around $f_9$. Secondly, these unexplained signals may be artefacts due to
amplitude and/or frequency variations. We recall from Sect.\ 4.1.1 that
although such temporal variability in the amplitude spectra is definitely
present, we cannot prove this hypothesis. In addition, the presence of a
``pure'' triplet with a $14.5\,\mu$Hz spacing in April 2007 is hard to
explain with amplitude/frequency variability only.

The period spacing between the normal modes of a pulsating white dwarf
star serves to obtain its mass if it allows a determination of the
asymptotic g-mode period separation. Following the discussion above, we
interpret $f_1$, $f_2$, $f_5$, $f_9$, $f_{13}$, $f_{16}$, $f_{18}$ and
$f_{19}$ as normal mode frequencies with an azimuthal order $m=0$, but 
with different radial overtone $k$ and/or spherical degree $\ell$. The
assignment of $m$ follows from the multiplet structure around $f_9$,
$f_{13}$ and $f_{16}$ (in case they are $\ell=1$ modes), but is assumed
for the others. However, if wrong, this assumption would not significantly
affect any mass determination, because the frequency difference of
consecutive radial overtones is significantly larger than the rotational
splitting.

In any case, our attempts to establish the asymptotic g-mode period
separation of EC~14012$-$1446 failed. There may be two possible causes:  
firstly, the effects of mode trapping (e.g.\ see Brassard et al.\ 1992)  
could be so strong that no regular period spacing can be found within the
few modes available. Secondly, not all of the modes we adopted may
have the same spherical degree $\ell$. A comparison of the periods of
these eight modes with the ones identified as $\ell=1$ by Kleinman et al.\
(1998) in G\,29-38 shows good agreement except for $f_{18}$, assuming the
stars are homologous. We conclude that an asteroseismic mass determination
for EC~14012$-$1446 must either await the detection of a more complete set
of normal modes or requires systematic matching with a model grid
(Metcalfe, Montgomery \& Kanaan 2004, Castanheira \& Kepler 2008).

With the independent frequencies and their amplitudes determined, we can 
compute the weighted mean pulsation period of EC~14012$-$1446 at the 
different observing epochs, adopting the definition by Mukadam et al.\ (2006):
\begin{equation}
$WMP$=\frac{\sum(P_iA_i)}{\sum(A_i)},
\end{equation}
where WMP is the weighted mean period, the $P_i$ are the periods of the
individual pulsation modes and the $A_i$ are their respective amplitudes.  
We also computed the rms scatter of each individual data point in our
light curves and the square root of the total power $(\sum(A_i)^2)^{1/2}$
as indications of the pulsation power of the star. The first measure makes
the (justified) assumption that the intrinsic variations in our light
curves dominate the measurement noise and it reflects the total 
variability. The second measure requires knowledge of the independent 
mode frequencies, but allows a direct comparison with the work by Mukadam 
et al.\ (2006).

\begin{table}
\caption[]{Changes in the light curves of EC~14012$-$1446.}
\begin{center}
\begin{tabular}{lccc}
\hline
Month/Year & WMP & $(\sum(A_i)^2)^{1/2}$ & rms scatter \\
& (s) & (mmag) & (mmag) \\
\hline
June 1994 & 598.4 & 66.9 & $---$\\
April 2004 & 658.9 & 59.9 & 48.9\\
June 2004 & 663.6 & 64.9 & 53.6\\
May 2005 & 684.4 & 72.2 & 60.3\\
April 2007 & 692.0 & 85.7 & 71.2\\
\hline
\end{tabular}
\end{center}
\end{table}

The results of these calculations are given in Table 6. We also computed
the WMP and the square root of the total power from the data by Stobie et
al.\ (1995) for completeness. As these authors quoted nightly amplitudes 
for their independent frequencies, we adopted the mean of these amplitudes 
to obtain the WMP and $(\sum(A_i)^2)^{1/2}$ for this observing period.

Table 6 shows a clear trend: the mean period of the variability increased
from one observing epoch to the next (a fact that can already be seen by
eye in Fig.\ 2). The total light amplitude increased during our
observations, but was at an intermediate level when Stobie et al.\ (1995)
observed the star. As with the apparent frequency changes of some
individual modes (Sect.\ 4.1) we caution against interpreting the change
in the WMP as a monotonous trend in time: the gaps between our observing
runs were considerably larger than the observing runs themselves. Because
we have already found amplitude and frequency changes within one observing
run, it is quite possible that the star altered its behaviour on time
scales not sampled by us.

If we compare these results with the sample of white dwarfs analysed by
Mukadam et al.\ (2006), we find that EC~14012$-$1446 is among the
highest-amplitude ZZ Ceti stars and that it is located close to the centre
of the instability strip, consistent with spectroscopic results (Fontaine
et al.\ 2003, Bergeron et al.\ 2004). The high amplitudes let us speculate
that most pulsation modes are indeed $\ell=1$, suffering the smallest
geometrical cancellation. The temporal variation of the WMP we observed is
not unusual compared to other ZZ Ceti stars (Mukadam et al.\ 2007).

\subsection{The second-order combination frequencies}

Most of the second-order combination signals have amplitudes that scale
with the product of the amplitudes of their parent modes. These are
consistent with an interpretation in terms of light-curve distortions.
Consequently, we can try to determine some convection parameters with
light-curve fitting following Handler, Romero-Colmenero \& Montgomery
(2002), based on Wu's (2001) analytical expressions for the combination
signals' amplitude and phases (also see Yeates et al.\ 2005).

Using the data from April 2004, assuming all parent modes are $\ell=1$ and
only using the mode triplets, we find a response parameter
$2\beta+\gamma=9.8\pm0.3$, convective thermal time $\tau_c=135\pm10$\,s,
and an inclination angle $\theta=29.5\pm1.5$\degr ~(see Wu 2001 for the
definitions of these parameters) when fitting the light curve. For 2005
however, no satisfactory fit could be obtained.


As noted before, our best candidate for a resonantly coupled mode is the
$2196\, \mu$Hz signal in the 2004 data, as its relative amplitude is by
far the highest, and since it varied in accordance with the amplitudes of
its parents $f_1$ and $f_7$. However, the interpretation of this $2196\,
\mu$Hz signal may be a chicken-and-egg problem: in June 2004, this signal
had about 70 per cent of the amplitude of its parent mode $f_1$. Given
that this number is close to unity and that different spherical degrees,
and thus different geometrical cancellation, of the parent and the
combination frequencies, may come into play, some doubt as to which signal
is indeed the parent and which one is the combination remains. In 
addition, a $2196\,\mu$Hz signal fits much better into the sequence of the
observed mode frequencies of EC~14012$-$1446 than $f_1=821\,\mu$Hz does.

\subsection{The third-order combination frequencies}

The large number of third-order combination frequencies depicted in Fig.\
6 follows a simple structure: all these groups of peaks are separated from
the dominant modes by the frequency differences between the three
strongest modes and their rotationally split components. This means we see
a pattern fairly regularly spaced in frequency, which, to first order, is
also a pattern regularly spaced in period.

If those third-order combinations are not correctly identified, but are
misinterpreted as independent frequencies, an incorrect determination of
the period spacing of the normal modes may therefore result. As an
experiment, we determined the mean period spacing of all independent and
third-order combination frequencies detected between 950 and 2550~$\mu$Hz
in the May 2005 data. We arrived at a formally significant mean period
spacing of 17.7~s. Some ``radial overtones'' in this scenario would be
missing, but no integer multiple of 17.7~s would give a reasonable fit.
Incidentally, this completely incorrect period spacing is close to the
true period spacing of the massive ZZ Ceti star BPM 37093 (Kanaan et al.\
2005).  Hence, EC~14012$-$1446 could be misinterpreted as having high
mass, inconsistent with spectroscopic results (Fontaine et al.\ 2003). It
is therefore essential that combination frequencies are correctly
recognized in the amplitude spectra of pulsating white dwarf stars.

Vuille et al.\ (2000) pointed out a similar problem in their study of the
prototype pulsating DB white dwarf star GD 358. However, there are some
differences between our results and those by Vuille et al.\ (2000): the
third-order combination frequencies revealed in their work only manifested
themselves as ``odd'' peaks lying close to or within rotationally split
multiplets, and these authors could explain all of these ``odd'' peaks
with combination frequencies. In our case, the third-order combinations
create apparent additional frequency multiplets, and we found signals
within the rotationally split normal mode multiplets that we were unable
to explain by combination frequencies; we suspect they are artefacts
caused by amplitude/frequency variations during the measurements.

\section{Conclusions}

We have acquired more than 200 hours of time-resolved photometry of the 
ZZ Ceti star EC~14012$-$1446. The star turned out to have a complex 
pulsation spectrum, that we were able to understand to some extent, but 
not to our full satisfaction.

We detected eight normal modes of the star with different radial
overtones. Not all of them must necessarily have the same spherical degree
$\ell$. Some of the normal modes are arranged in groups, in most cases
showing separations of $\approx 10\, \mu$Hz. We interpreted that as the
sign of rotational m-mode splitting, and inferred the rotation period of
the star ($0.61\pm0.03$\,d), assuming modes of $\ell=1$. 

We were unable to determine a regular period spacing between the pulsation 
modes of EC~14012$-$1446, and therefore cannot provide a mass estimate. 
The reason may be mode trapping or that we see a mixture of modes with 
different spherical degree $\ell$.

The pulsation spectrum of the star changes over time. We have measured
changes of $\pm 8$\% in the weighted mean period and of $\pm 20$\% in the
square root of the total pulsation power. Our data have insufficient
sampling and time base to determine the time scale of the variations in
the normal modes. However, we found systematics in the temporal changes of
the pulsation spectrum: the higher the radial overtone of a mode, the more
pronounced the variations are and amplitude and frequency changes go hand
in hand. The dependence on radial overtone is easy to understand: since
the kinetic energy of the g modes in a pulsating white dwarf star becomes
weighted towards the outer regions with increasing radial overtone, less
energy is required to disturb a high-overtone mode.

Most of the second-order combination frequencies have amplitudes
consistent with the hypothesis that they originate from light curve
distortions of the star. However, some could be resonantly excited modes
because they have relative amplitudes much higher than the other
second-order combinations. We even found one case where it was no longer
clear which was the parent mode and which was the combination, although
this could partly be an effect of geometrical cancellation.

The amplitudes of the second-order combinations varied in time in
accordance with their parent modes. The exceptions to this rule are a
number of combinations that involve the mode $f_9$. This mode is also
special in the sense that it shows complicated ``multiplet'' structure in
one of our data sets, that we tentatively attributed to the effects of
amplitude and frequency variations during the observations.

Finally, we showed how a misinterpretation of third-order combination
frequencies of the form $f_A+f_B-f_C$ as normal modes, can lead to a
completely nonsensical determination of the stellar mass. Combination
frequencies intruding into the domain of the normal modes in the Fourier
spectra of the light curves must be carefully searched for, and eliminated
from an asteroseismic analysis.

The ZZ Ceti star EC~14012$-$1446 remains an attractive target for further
study, as we could only partly understand its pulsations. The high
pulsation amplitudes of the star may facilitate the application of
Montgomery's (2005) method to determine convection parameters and mode
identifications from high signal-to-noise light curves. The detection of
more normal modes of pulsation and the correct identifications of at least
some of them may permit a full asteroseismic analysis. A reliable
determination of the time scales of the amplitude and frequency variations
of the star would be worthwhile to pin down their origin, but would
require a data set with a time base in excess of two weeks, excellent duty
cycle at the same time, as well as a co-operating target. Under such
circumstances, precision asteroseismology of a ZZ Ceti star may be
feasible.

\section*{ACKNOWLEDGEMENTS}

This work has been supported by the Austrian Fonds zur F\"orderung der
wissenschaftlichen Forschung under grant P18339-N08, Mt.\ Cuba Observatory
and the Delaware Asteroseismic Research Center. This material is based
upon work supported by the National Science Foundation under Grant No.\
AST 0205902 to the Florida Institute of Technology. GH thanks Barbara
Castanheira Endl and Victoria Antoci for commenting on a draft version of
this paper.

\bsp

\end{document}